\documentclass[11pt, twoside, table]{predoc2}

\usepackage[utf8]{inputenc}
\usepackage[T1]{fontenc}
\usepackage{newtxtext} 
\usepackage[scaled=.95]{cabin} 
\usepackage[varqu,varl]{inconsolata} 
\usepackage{enumitem}
\usepackage{tabularx}
\usepackage[varg]{newtxmath}
\usepackage{changepage}
\usepackage{hyperref}
\usepackage{url}            
\usepackage{graphicx}
\usepackage{subcaption} 
\usepackage{float}
\usepackage{booktabs}       
\usepackage{nicefrac}       
\usepackage{microtype}      
\usepackage{xcolor}         
\usepackage{bm}
\usepackage{multirow}
\usepackage{moreverb,url}

\hypersetup{
  urlcolor = red,
  colorlinks = true
}
\usepackage{amsmath,amssymb,amsfonts}
\newtheorem{theorem}{Theorem}[section] 
\newtheorem {lemme}{Lemma} 
\newtheorem {definition}{Definition}[section] 

\title{High-Dimensional Data with Measurement Error}
\author[1]{Herman Tesso}
\affil[1]{AIMS South Africa, Aalto University, Finland \hspace{3cm}  
  \texttt{herman.tesso@aalto.fi}}
\author[2]{Georges Nguefack-Tsague}
\affil[2]{University of Yaounde 1, Cameroon \hspace{4.4cm} \texttt{nguefacktsague@gmail.com}}

\makeatletter
\def\runauthor{Tesso and Nguefack-Tsague}
\def\runtitle{High-dimensional regression with measurement error}
\makeatother

\begin{document}

\maketitle
\begin{abstract}
In many important statistical analyses, the number of covariates $p$ often exceeds the data size $n$, a regime commonly referred to as high-dimensional. While considerable progress has been made in high-dimensional regression under the assumption of error-free covariates, real-world data frequently involve noisy or corrupted measurements. When left unaddressed, measurement errors can silently distort the analysis and mislead the conclusions. This paper reviews and evaluates some advisable statistical inference methods for high-dimensional regression in the presence of mismeasured covariates. We discuss four penalized regression methods--- ridge, lasso, Dantzig selector, and Elastic-net---alongside their measurement-error-corrected variants, and conduct a comparative study under linear additive and uncorrelated measurement error models.
Through simulation studies and a real application to high-dimensional medical genetic data, we illustrate the methods studied, show that the choice of correction procedure is problem-specific, and provide practical recommendations to help practitioners navigate this methodological landscape.\\

\end{abstract}
\thispagestyle{empty}

\section{Introduction}

The past two decades have witnessed an abundance of large-scale data across scientific disciplines, driven by rapid advances in automated data collection and measurement technologies. In genetics, for instance, a new analytical paradigm has emerged in which the number of variables $p$ far exceeds the number of observations $n$---a regime commonly referred to as high-dimensional. High-dimensional data analysis has grown considerably in both theoretical development and practical application, and a wide range of methods have been proposed for statistical modeling and inference in this context. Penalized regression methods such as ridge regression \citep{nref8}, lasso \citep{nref9}, and the Dantzig selector (DS) \citep{nref12} discussed in this paper collectively form a well-established toolkit for high-dimensional regression under ideal conditions, where covariates are error-free. 

In practice, however, exact measurement of variables of interest is rarely achievable. Researchers must therefore work with error-prone proxies of the true underlying quantities. Measurement errors may arise from instrument imprecision, sampling variability, or data collection limitations. Although the consequences of ignoring measurement error have been recognized for some time, they range from negligible to severe depending on the setting. Despite this, the majority of researchers do not correct for measurement error — even when they are aware of its presence — partly because the auxiliary information required for correction is often unavailable.
There are at least three reasons why measurement error should not be ignored. First, it can induce bias in parameter estimation \citep{nref10}. Second, it can interfere with variable selection \citep{nref17}. Third, it can reduce statistical power \citep{nref11}, making it harder to detect true relationships among variables. Crucially, analyzing the bias of naive estimators often naturally suggests an appropriate correction procedure.
This has motivated a growing body of work on measurement-error-corrected penalized regression. The present paper reviews these methods, examining their theoretical foundations, practical implementations, and comparative performance.

The paper is organized as follows. We begin by discussing the challenges of high-dimensional data and the standard linear regression setup, before laying out the theoretical framework of measurement error in regression and its inferential consequences. We then extend this framework to the high-dimensional setting, describing measurement-error-corrected variants of ridge, lasso, and the Dantzig selector. The methods are subsequently evaluated and compared through simulation studies and a real application to high-dimensional medical genetic data, followed by a discussion and concluding remarks.

\section{Challenge with high-dimensional data}
\label{sec:1}

High-dimensional data should not be mistaken for \emph{big data}, which is a term for massive datasets characterized by a large volume of information, high velocity of data generation, and substantial variety in data types---referred to as the ''\textit{the three Vs}'' \citep{sagiroglu2013big}. In fact, not all big data is necessarily high-dimensional. The challenges addressed here arise specifically
from an unfavorable feature-to-sample ratio, and are present irrespective of the absolute
size of the dataset. In medical research, such settings are commonplace. 
For instance, one may wish to predict a
patient's blood pressure using age, BMI, and cholesterol level, alongside
genotype information from 200{,}000 single nucleotide polymorphisms measured
across only 100 patients. Similarly, estimating the probability of cancer
progression may involve gene expression data from 20{,}000 genes combined
with general health indicators (age, weight, BMI, blood pressure) and 
tumor teristics (stage, localized spread, biomarker results). In both cases,
the number of variables (predictors) $p$ far exceeds the number of observations $n$,
placing them squarely in the high-dimensional regime.

This having more features than observations introduces a range of difficulties that
standard statistical methods are not equipped to handle. The following development uses ordinary least squares regression (OLS) as a concrete
illustration of these failures, but the same issues apply, in principle, to
logistic regression, linear discriminant analysis, and other classical approaches.

\subsection{OLS failure in high-dimensional setting}
\label{subsec:ols-failure}
To motivate the need for specialized methods, it is instructive to examine
precisely how ordinary OLS degenerates when the number of predictors $p$ is much larger than the data size $n$. Consider the standard multiple linear regression model
\begin{equation}
    y = X\beta + \varepsilon,
    \label{eq:linreg}
\end{equation}
where $y \in \mathbb{R}^n$ is the response variable, $X\in \mathbb{R}^{n \times p}$
is the design matrix, $\beta \in \mathbb{R}^p$ is the unknown coefficient vector,
and $\varepsilon \sim \mathcal{N}(0,\, \sigma^2 I_n)$ is the noise. The OLS estimator minimizes the residual sum of squares,
\begin{equation}
    \hat{\beta}^{\text{LS}} = \operatorname{arg\,min}_{\beta}\,
    \|y - X\beta\|_{\ell_2}^2
    = \bigl(X^{\top}X\bigr)^{-1}X^{\top}y,
    \label{eq:ols}
\end{equation}
and the corresponding covariance and residual variance estimate are given by
\begin{equation*}
    \text{Cov}(\hat{\beta}^{\text{LS}}) = \sigma^2(X^{\top}X)^{-1},\ \ \ \ s^2 = \frac{1}{n-p-1}\sum_{i=1}^{n}(y_i - \hat{y}_i)^2,
\end{equation*}
provided the Gram matrix $X^{\top}X$ is invertible---a condition
that requires $X$ to have full column rank $p$. 

When $p > n$, however,
$X$ has at most rank $n < p$, so $X^{\top}X$ is necessarily
singular (non-invertible) and the OLS estimator in Eq.~\eqref{eq:ols} becomes ill-defined. Therefore, infinitely many solutions satisfy the normal equations $X^{\top}X\beta=X^{\top}y$, and $s^2$ is no longer well-defined since $n - p - 1 < 0$. 

Even if $p$ is only moderately large relative to $n$, the matrix $X^{\top}X$ may be near-singular (ill-conditioned) due to potential multicollinearity among predictors, which becomes increasingly likely as $p$ grows larger. 
The practical consequences of near-singularity can be quantified through the
mean squared error (MSE), $ MSE({\hat{\beta}}^{\text{LS}})=\mathbb{E}\bigl[\|{\hat{\beta}}^{\text{LS}} - \beta\|_{\ell_2}^2\bigr]$, of the OLS estimator which decomposes into bias squared plus variance. Writing the eigen-decomposition
$X^{\top}X = V\,\mathrm{diag}(\lambda_1, \ldots, \lambda_p)\,V^{\top}$, one obtains
\begin{equation}
    \mathbb{E}\bigl[\|{\hat{\beta}}^{\text{LS}} - \beta\|_{\ell_2}^2\bigr]
    = \sigma^2\, \mathrm{tr}\bigl[(X^{\top}X)^{-1}\bigr]
    = \sigma^2 \sum_{i} \frac{1}{\lambda_i}.
    \label{eq:mse-ols}
\end{equation}
Equation~\eqref{eq:mse-ols} reveals a critical vulnerability: whenever any
eigenvalue $\lambda_i$ is small, the corresponding term $1/\lambda_i$ dramatically
inflates the MSE between the least squares estimate and the true parameter. The estimates are formally valid but numerically unstable, with inflated variance. 

For example, if $\lambda_i = 10^{-5}$, then
the contribution of that single eigenvalue alone inflates
$\hat{\beta}^{\text{LS}}$ above the true $\beta$ by a
factor of $10^5 \sigma^2$. 

\begin{lemme}
\label{lem:1}
	An $n\times n$ ill-conditioned or near singular matrix has at least one of its eigenvalues close to zero, and the eigenvalues of the inverse tend to be very large.
\end{lemme}	
Beyond algebraic and numerical failures, common issues such as the \emph{bias--variance trade-off} and \emph{overfitting}, well understood in the low-dimensional regime where $p < n$, are substantially amplified when $p \gg n$ and cannot be managed with standard diagnostics or model selection criteria alone. 

These challenges motivate the use of appropriate regularization techniques which we discussed next.

\section{Inference methods for high-dimensional regression analysis}
\label{s2}
\subsection{Ridge regression}

\label{rre}
Ridge regression addresses multicollinearity-induced instability in OLS by trading bias for variance reduction.
Although the Gauss-Markov theorem guarantees that the OLS estimator has minimum variance among all unbiased estimators, it does not ensure minimum MSE. Ridge regression deliberately introduces bias by adding a constraint (quadratic or $\ell_2$-penalty) on the coefficient magnitudes, substantially reducing variance and often achieving lower MSE. This regularization tend to stabilizes estimation even when \(X^TX\) is ill-conditioned (near-singular).

For any estimator $\beta$, the least squares criterion $\mathcal{Q}(\beta)= \| y-X\beta \|_{\ell_2}^{2}$ can be rewritten as its minimum, reached at $\hat{\beta}^{LS}$ plus a quadratic form in $\beta$:
$$ \mathcal{Q}(\beta)=\underbrace{\| y-X\hat{\beta}^{LS} \|_{\ell_2}^{2}}_{\mathcal{Q}_{min}}+\underbrace{( \hat{\beta}^{LS}-\beta)^{\top}X^{\top}X( \hat{\beta}^{LS}-\beta)}_{\phi(\beta)}.$$

The optimization problem in ridge regression can be then be stated as  minimizing
$$ \{\| \beta \|_{\ell_2}^{2}\}\ \ s.t\ \ ( \hat{\beta}^{LS}-\beta)^{\top}X^{\top}X( \hat{\beta}^{LS}-\beta)=\phi_{0}$$
for some constant $\phi_{0}$.
The enforced constraint $\phi_{0} \ $ guarantees a relatively small residual sum of squares $\mathcal{Q}(\beta)$ when compared to its minimum $\mathcal{Q}_{min}$. As a Lagrangian problem, this is equivalent to
\begin{equation}
     \operatorname{arg\,min}_{\beta}\,\left\{\| \beta \|_{\ell_2}^{2} + \frac{1}{k}\left[( \hat{\beta}^{LS}-\beta)^{\top}X^{\top}X( \hat{\beta}^{LS}-\beta)-\phi_{0}\right]\right\},
     \label{f12}
 \end{equation}
 where $\frac{1}{k},\ (k>0\ )$, is the multiplier chosen to satisfy the constraint.
The numerical solution of this problem corresponding to the ridge regression estimator of $\beta$ is
\begin{equation}
\hat{\beta}^{R}=(X^{\top}X+k\mathbb{I}_{p})^{-1}X^{\top}y.
\label{f11}
\end{equation}
Setting \( k = 0 \) recovers the (unregularized) OLS estimator. Let $\lambda_{max} = \lambda_{1} \geq \lambda_{2} \geq...\geq \lambda_{p}=\lambda_{min}$ denote the eigenvalues of $X^{\top}X$ in decreasing order. Then the eigenvalues of $Z=(X^{t}X+k\mathbb{I}_{p})^{-1}$ are $\frac{\lambda_{j}}{\lambda_{j} + k},\ j=1,...,p,$ and the mean square error, $MSE(\hat{\beta}^{R},k)=\mathbb{E}\bigl[\|{\hat{\beta}}^{\text{R}} - \beta\|_{\ell_2}^2\bigr],$ of ridge regression estimator is given by
\begin{equation}
    MSE(\hat{\beta}^{R},k)=k^{2}\beta^{\top}(X^{\top}X+k\mathbb{I})^{-2}\beta +\sigma^{2}\sum_{j}\frac{\lambda_{j}}{(\lambda_{j}+k)^{2}}.
\label{f17}
\end{equation}
The constant $k$ reflects the amount of bias increased and the variance reduced. \cite{nref8} demonstrate that there always exists a value $k > 0$ such that,
$$MSE(\hat{\beta}^{R},k)<MSE(\hat{\beta}^{R},0)=MSE(\hat{\beta}^{LS}).$$


\subsection{Lasso regression}
\label{subsec:lasso}

The lasso (Least Absolute Shrinkage and Selection Operator) is a shrinkage
method closely related to ridge regression, but with one additional feature: unlike ridge, it performs automatic variable selection by shrinking a subset of the estimated coefficients exactly to zero, under the assumption that the true coefficient vector is sparse.
The $\ell_2$-penalty in ~\eqref{f12} shrinks coefficients
toward zero but never sets them exactly to zero, which makes interpretability of the model very difficult when $p$ is very large. Classical variable selection procedures such as best subset, forward stepwise, and backward stepwise selection,\footnote{These methods search for a parsimonious model in the
classical low-dimensional regime, where $n > p$.} do produce sparse
models, but are computationally infeasible in high dimensions. The lasso
resolves both shortcomings simultaneously: it delivers sparse, interpretable
models through a convex optimization problem that scales efficiently with $p$.
Specifically, the lasso estimator of $\beta$ solves
\begin{equation}
    \underset{\beta \in \mathbb{R}^{p}}{\operatorname{arg\,min}}
    \left\{ \| y - X\beta \|_{\ell_2}^{2} + \lambda \| \beta \|_{\ell_1} \right\},
    \label{f18}
\end{equation}
where the $\ell_1$-penalty $\lambda \|\beta\|_{\ell_1} = \lambda \sum_{j=1}^p
|\beta_j|$ is the key component that induces sparsity: unlike the
$\ell_2$-penalty, it imposes a diamond-shaped constraint region whose
corners lie on the coordinate axes, making it geometrically favourable
for solutions in which some coefficients are exactly zero.

\subsubsection{Theoretical properties of the lasso.}
\label{subsec:lasso-theory}

A central assumption underlying the lasso is \emph{sparsity}: only a small
number $s \ll p$ of covariates truly influence the outcome. Formally, let
$S = \{ j : \beta_j \neq 0 \}$ denote the index set of non-zero coefficients,
with $s = \mathrm{card}(S)$. For any $\lambda \geq 0$, the estimated active
set is $\hat{S}(\lambda) = \{ j : \hat{\beta}_j(\lambda) \neq 0 \}$. We
partition the covariates as $X = (X_S,\, X_{S^c})$, where $X_S \in
\mathbb{R}^{n \times s}$ collects the $s$ more relevant predictors and $X_{S^c}
\in \mathbb{R}^{n \times (p-s)}$ the remaining $(p-s)$, and define
\begin{equation}
    \beta_{j,S} = \beta_j \boldsymbol{1}_{\{j \in S\}},
    \qquad
    \beta_{j,S^c} = \beta_j \boldsymbol{1}_{\{j \notin S\}},
    \label{f33}
\end{equation}
so that $\beta = \beta_S + \beta_{S^c}$, with $\beta_S$ vanishing outside
$S$ and $\beta_{S^c}$ vanishing on $S$. 
Denote $\hat{\Sigma}_{X} = X^{\top} X / n$.

\paragraph{Estimation error bounds.}
A key condition governing the lasso performance is the \emph{compatibility
condition} \citep{nref14}.

\begin{definition}[Compatibility condition]
The compatibility condition holds for the set $S$ with constant $\Phi_0 > 0$
if, for all $\beta \in \mathbb{R}^p$ satisfying $\|\beta_{S^c}\|_1 \leq
3\|\beta_S\|_1$,
\begin{equation}
    \|\beta_S\|_1^2
    \;\leq\;
    \frac{s\, (\beta^{\top} \hat{\Sigma}_{X}  \beta)}{\Phi_0^2}.
    \label{f34}
\end{equation}
\end{definition}

\noindent
Intuitively, this condition ensures that the design matrix $X$ does not
spread the signal of the active predictors too diffusely across the inactive
ones, so that the lasso can reliably distinguish the two groups. Under this condition, the following finite-sample error bound holds
\citep[Theorem~6.1, p.~107]{nref14}: for $\lambda \geq 2\lambda_0$,
\begin{equation}
    \frac{1}{n}
    \underbrace{\| X(\hat{\beta}^{\mathrm{Lasso}} - \beta) \|_{\ell_2}^2}
    _{\text{prediction error}}
    +\;
    \lambda\,
    \underbrace{\| \hat{\beta}^{\mathrm{Lasso}} - \beta \|_{\ell_1}}
    _{L_1\text{-error}}
    \;\leq\;
    \frac{4\lambda^2 s}{\Phi_0^2}.
    \label{f35}
\end{equation}
This bound shows that both the prediction error and the $L_1$ estimation
error decrease as $\lambda$ decreases, and worsen as the number of active
predictors $s$ grows. That is, the lasso recovers the true
coefficient vector more accurately when the true model is sparser.
Under these compatibility assumptions, \citet{knight2000asymptotics} further show
that, for $\lambda$ of order $\sqrt{\log(p)/n}$,
\begin{equation}
    \| \hat{\beta}^{\mathrm{Lasso}} - \beta \|_{\ell_1}
    \xrightarrow{\;\mathbb{P}\;} 0
    \quad \text{and} \quad
    \| \hat{\beta}^{\mathrm{Lasso}} - \beta \|_{\ell_2}
    \xrightarrow{\;\mathbb{P}\;} 0
    \quad \text{as } n \to \infty,
    \label{f39}
\end{equation}
establishing consistency of the lasso estimator in the $\ell_1$ and
$\ell_2$ senses.

\paragraph{Model selection consistency.}
Beyond accurate estimation of $\beta,$ one may require the lasso to recover
the correct sparsity pattern, that is, to identify the true active set $S$
exactly. These are distinct requirements: an estimator can be consistent
for $\beta$ without consistently selecting the correct model, and vice
versa. Formally, parameter estimation consistency requires
$\hat{\beta}^{\mathrm{Lasso}} - \beta \xrightarrow{\mathbb{P}} 0$,
whereas model selection consistency requires
$\mathbb{P}(\hat{S} = S) \to 1$ as $n \to \infty$. Ideally, one seeks
an estimator that satisfies both.

Model selection consistency very much depends on the choice of
regularization parameter $\lambda$: one may ask either whether a fixed,
deterministic $\lambda$ achieves consistent selection, or whether, for
each realization of the data, there exists a $\lambda$ that recovers $S$
exactly. \citet{nref15} demonstrate that both questions are governed by the ''\emph{irrepresentable condition}''.

\begin{definition}[Irrepresentable condition]
The irrepresentable condition holds for the set $S$ if there exists a
constant $\theta \in [0, 1)$ such that
\begin{equation}
    \left\|
        \hat{\Sigma}_{X}(S^c, S)\, \hat{\Sigma}_{X}(S, S)^{-1}\, \mathrm{sign}(\beta_S)
    \right\|_\infty
    \leq \theta.
    \label{f40}
\end{equation}
\end{definition}

\noindent
In essence, this condition requires that the inactive predictors
$X_{S^c}$ are not too correlated with the active predictors $X_S$,
so that the Lasso penalty can suppress the former without distorting
the latter. The following theorem formalises its role.

\begin{theorem}[\citealt{nref15}]
The irrepresentable condition~\eqref{f40} is a sufficient and essentially
necessary condition for the lasso to select only variables in the active set $S$.
\end{theorem}

\subsection{Elastic-net regression}
\label{subsec:elasticnet}

Elastic-net regression combines the $\ell_1$-penalty of the lasso with the
$\ell_2$-penalty of ridge regression into a single, unified framework. This
combination is motivated by two well-known limitations of the lasso: it
tends to select at most $n$ variables when $p \gg n$, and it struggles with
groups of highly correlated predictors, typically selecting one arbitrarily
while discarding the rest. By incorporating a ridge component, the Elastic-net
overcomes both shortcomings while retaining the sparsity-inducing property
of the lasso.

In its most general form, the Elastic-net estimator solves
\begin{equation}
    \underset{\beta \in \mathbb{R}^p}{\mathrm{argmin}}
    \left\{
        \| y - X\beta \|_{\ell_2}^2
        + \lambda_1 \| \beta \|_{\ell_1}
        + \lambda_2 \| \beta \|_{\ell_2}^2
    \right\},
    \label{eq:enet-general}
\end{equation}
where $\lambda_1 \geq 0$ and $\lambda_2 \geq 0$ control the contribution
of the lasso and ridge penalties, respectively. In practice, the
two-parameter formulation is more conveniently reparametrized using a
single regularization parameter $\lambda \geq 0$ and a mixing parameter (weight) 
$\alpha \in [0, 1],\;   \hat{\beta}^{E}(\lambda, \alpha)
    $:
\begin{equation*}
    \underset{\beta \in \mathbb{R}^p}{\mathrm{argmin}}
    \left\{
        \| y - X\beta \|_{\ell_2}^2
        + \lambda
        \bigl(
            \alpha \| \beta \|_{\ell_1}
            + (1 - \alpha) \| \beta \|_{\ell_2}^2
        \bigr)
    \right\},
 ;,\end{equation*}
where $\alpha$ interpolates between pure ridge ($\alpha = 0$) and pure
Lasso ($\alpha = 1$). This parametrization makes the relationship between
the three methods transparent:
\begin{equation*}
    \hat{\beta}^{E}(\lambda, 1) = \hat{\beta}^{\mathrm{Lasso}}(\lambda),
    \ \
    \hat{\beta}^{E}(\lambda, 0) = \hat{\beta}^{R}(\lambda),
    \ \
    \hat{\beta}^{E}(0, \alpha)  = \hat{\beta}^{LS}.
\end{equation*}
For $\alpha \in (0, 1)$ and $\lambda > 0$, the Elastic-net provides a hybrid estimator that simultaneously shrinks coefficients
(via the $\ell_2$ term) and sets some of them to zero (via the $\ell_1$
term). This offers an effective compromise in situations where the competing demands of parsimony and predictive stability cannot be adequately addressed by either penalty alone.

\subsection{Dantzig selector}
\label{subsec:dantzig}

The Dantzig selector (DS), proposed by \citet{nref12}, is another
$\ell_1$-penalised method for high-dimensional regression. Like the lasso,
it produces sparse solutions by setting a subset of coefficients to zero.
The key conceptual difference lies in how the two methods control the fit:
whereas the lasso minimizes the squared error loss subject to an $\ell_1$
constraint on $\beta$, the Dantzig selector instead controls the
$\ell_\infty$ norm of the correlation between the residuals and the
predictors. Specifically, it is defined as the solution to
\begin{equation}
    \underset{\beta \in \mathbb{R}^p}{\mathrm{minimize}}
    \; \| \beta \|_{\ell_1}
    \quad \text{subject to} \quad
    \| X^{\top} (y - X\beta) \|_\infty \leq \lambda\sigma,
    \label{f41}
\end{equation}
where $\lambda > 0$ is a tuning parameter, $\sigma$ is the noise level,
and $r = y - X\beta$ denotes the residual vector. The constraint requires
that, for each $j \in \{1, \ldots, p\}$,
$| (X^{\top} r)_j | \leq \lambda \sigma$,
ensuring that no predictor remains substantially correlated with the
residuals — a natural notion of a well-fitted model. When $\lambda_p$ is
chosen of order $\sqrt{\log p / n}$, this constraint is satisfied with
high probability under standard assumptions on the noise.

Despite their different motivations, the Dantzig selector and the lasso are
closely connected. Both perform variable selection by shrinking some
coefficients exactly to zero, and under certain regularity conditions, they
can yield the same solution path \citep{james2009dasso}. The fundamental
distinction is conceptual: the lasso arises from a penalized likelihood or
objective function, whereas the Dantzig selector is derived directly from
the score equation of the linear model, making it closer in spirit to an
estimating-equation approach. Formal theoretical guarantees for the Dantzig
selector — covering both estimation accuracy and model selection consistency
— are established in \citet[][Theorems~1.1 and~1.2]{nref12}.

\subsection{Cross-validation for hyperparameter tuning}
\label{subsec:cv}

All penalized regression methods discussed so far --- ridge, lasso, and
Elastic-net --- depend on a regularization parameter $\lambda$ (or $k$) that
controls the strength of the penalty. Since the optimal value of $\lambda$
is unknown in practice, it must be estimated from the data. Cross-validation (CV)
is the standard approach for doing so.

In $K$-fold cross-validation, the data are partitioned into $K$ roughly
equal subsets, or folds. For each fold $k = 1, \ldots, K$, a model
is fitted on the remaining $K - 1$ folds and its prediction error is
evaluated on fold $k$, which serves as a held-out validation set. This
process is repeated for each fold in turn, so that every observation
contributes to the validation exactly once. The optimal $\lambda$ is then
selected as the value that minimizes the total cross-validation error,
averaged across all folds:
\begin{equation}
    \hat{\lambda}
    = \underset{\lambda}{\mathrm{argmin}}
    \left\{
        \frac{1}{K} \sum_{k=1}^{K}
        \frac{1}{n_k}
        \| y_k - X_k \hat{\beta}_k(\lambda) \|^2
    \right\},
    \label{f42}
\end{equation}
where $(X_k, y_k)$ denotes the design matrix and response vector for the
$k$-th fold, $n_k$ is the number of observations in that fold, and
$\hat{\beta}_k(\lambda)$ is the coefficient estimate obtained by fitting
the model on all folds except the $k$-th. In practice, $K = 5$ or $K = 10$ are common choices, balancing computational efficiency with variability of the CV error estimate. The limiting case $K = n$, where each observation forms its own fold, corresponds to leave-one-out cross-validation, which minimizes variance in the error estimate but incurs the highest computational cost.

\section{Measurement error in regression theory}
\label{s3}

Measurement error arises whenever one or more variables entering a model
cannot be observed exactly. The most common sources are sampling variability
and instrument imprecision. Throughout this paper, the true but unobservable
covariate is denoted $X$, while its error-prone observed surrogate is denoted
$W$. When both quantities are categorical, this phenomenon is referred to as
misclassification. In this section, we introduce a formal model for measurement error, characterize the consequences of ignoring it, and describe available correction methods.

\subsection{Model formulation}
\label{subsec:me-model}

A fundamental assumption of linear regression is that covariates are observed without error. In the presence of measurement error, standard inferential tools become misleading, producing biased and inconsistent estimates \citep{nref2, nref3}. We illustrate by augmenting the linear regression model of~\eqref{eq:linreg} with an additive error model structure:
\begin{equation}
    y = X\beta + \varepsilon,
    \qquad
    W = X + U.
    \label{f46}
\end{equation}
This is equivalent to $y_{i}=\beta^{\top}X_{i}+\epsilon_{i},\ W_{i}=X_{i}+U_{i}, $  where, for each observation $i = 1, \ldots, n$, the vectors
$X_i = (X_{i1}, \ldots, X_{ip})^{\top}$,
$W_i = (W_{i1}, \ldots, W_{ip})^{\top}$, and
$U_i = (U_{i1}, \ldots, U_{ip})^{\top}$
collect the true covariates, their observed surrogates, and the measurement
errors, respectively. For notational simplicity, the intercept is set to $\beta_0 = 0$.

The true design matrix $X$ is latent (unobservable). Instead, we observe the
error-contaminated matrix $W = X + U$, where $U \in \mathbb{R}^{n \times p}$ is
a random noise matrix whose rows are independently normally distributed with
mean zero and covariance $\Sigma_U$ --- a known $p \times p$ matrix with
non-negative diagonal elements. If the $k$-th covariate is measured without
error, the corresponding column of $U$ is identically zero, so that
$\Sigma_{U(k,k)} = 0$. Throughout, $U$ and $\varepsilon$ are assumed to be
mutually independent.

Under the structural model defined by Eq.~\eqref{f46}, the joint vector
$(y_i, W_i^{\top})^{\top}$ follows a $(p+1)$-variate normal distribution with mean
$\mu = (\beta^{\top}\mu_X,\, \mu_X^{\top})^{\top}$ and block covariance matrix
\begin{equation}
    \Gamma
    = \begin{bmatrix}
        \sigma^2_y  & \Sigma_{yW} \\
        \Sigma_{Wy} & \Sigma_W
      \end{bmatrix}
    = \begin{bmatrix}
        \sigma^2 + \beta^{\top}\Sigma_X\beta & \beta^{\top}\Sigma_X      \\
        \Sigma_X\beta                   & \Sigma_X + \Sigma_U
      \end{bmatrix},
    \label{f48}
\end{equation}
where $\sigma^2_y = \sigma^2 + \beta^{\top}\Sigma_X\beta$ is the marginal
variance of $y$, and the off-diagonal blocks $\Sigma_{YW} = \beta^{\top}\Sigma_X$
and $\Sigma_{Wy} = \Sigma_X\beta$ reflect the dependence between the response
and the observed covariates induced by the shared true signal $X$. From the
joint distribution in Eq.~\eqref{f48}, the conditional distribution of $y_i$
given $W_i$ is linear:
\begin{equation}
    y_i \mid W_i\;=\; \gamma^{\top} W_i + \delta_i,
    \label{f49}
\end{equation}
where $\delta_1, \ldots, \delta_n$ are i.i.d.\ $\mathcal{N}(0,\,
\sigma^2_\delta)$. The conditional regression coefficient $\gamma$ and variance $\sigma^2_\delta$ are derived from Eq.~\eqref{f48} using the formula for the normal conditional distribution:
\begin{equation}
\begin{split}
        \gamma
   & = \bigl(\Sigma_X + \Sigma_U\bigr)^{-1}\Sigma_X\beta,
    \quad \text{and} \\
    \sigma^2_\delta
   & = \sigma^2 + \beta^{\top}\Sigma_X\beta
      - \gamma^{\top}\bigl(\Sigma_X + \Sigma_U\bigr)\gamma.
\end{split}
    \label{eq:gamma-sigma}
\end{equation}
Observe that $\gamma \neq \beta$ whenever $\Sigma_U \neq 0$. The observable
regression coefficient $\gamma$ is an attenuated (biased toward zero) version of the true
parameter $\beta$. Their relationship is made explicit by the following formula:
\begin{equation}
    \beta = \mathcal{K}_X^{-1}\gamma,
    \qquad
    \mathcal{K}_X = (\Sigma_X + \Sigma_U)^{-1}\Sigma_X,
    \label{f50}
\end{equation}
where $\mathcal{K}_X$ is the
$p \times p$ \emph{reliability matrix} \citep{nref4}. Each diagonal entry
of $\mathcal{K}_X$ lies in $[0,\,1]$ and quantifies the proportion of the
total observed variance in the corresponding column of $W$ that is
attributable to the true signal $X$; a value of one indicates perfect
measurement, while a value close to zero indicates severe contamination.

\subsection{Measurement error induces bias} \label{subsec:me-naive}

In practice, analysts may be unaware of measurement error or lack the information needed to correct for it. Common course of action proceed as though the observed covariates $W$ were error-free proxies for the unobserved true variables \(X\). This naive approach, however, produces estimators that inherit systematic bias from the unacknowledged contamination, as we shall see with OLS.

Following equation~\eqref{eq:ols}, applying standard OLS to the contaminated matrix $W$ gives estimates
\begin{equation*}
    \hat{\beta}_{\mathrm{naive}} = (W^{\top}W)^{-1}W^{\top}y,
    \quad
    \hat{\sigma}^{2}_{\mathrm{naive}} = \frac{1}{n-p-1}\sum_{i}(y_i - \hat{y}_{i})^{2},
    \label{f51}
\end{equation*}
where $\hat{y}_i = \hat{\beta}_{\mathrm{naive}}^{\top} W_i$. These coincide with the maximum likelihood estimates of $\gamma$ and $\sigma^2_\delta$ from the conditional model ~\eqref{f49}. Under the measurement error model~\eqref{f46}, the expected values of these estimators are
\begin{equation}
    \mathbb{E}\bigl[\hat{\beta}_{\mathrm{naive}}\bigr]
    = \mathcal{K}_X\beta = \gamma,
    \quad
    \mathbb{E}\bigl[\hat{\sigma}^2_{\mathrm{naive}}\bigr]
    = \sigma^2_\delta.
    \label{f52}
\end{equation}
This reveals that the naive OLS estimators are biased whenever $\Sigma_U \neq 0$. The \emph{reliability matrix} $\mathcal{K}_X$ characterizes this bias: since it is generally a full matrix, measurement error in even a single covariate contaminates the estimates of all coefficients, including those of error-free variables. With increasing number of mismeasured covariates, the structure of $\mathcal{K}_X$ becomes rather complex, and the effect of measurement error---propagation of bias throughout the coefficient vector---becomes difficult to describe.

\subsection{Correction methods}
\label{subsec:me-correction}
Correcting for measurement error generally requires auxiliary information beyond the primary sample, e.g., replicated measurements or instrumental variables. When such resources are unavailable, several alternative approaches exist, a number of which are described in \citep{nref10}. These include moment-based approaches, which exploit sample moment relationships; likelihood-based techniques, which specify and maximize a joint likelihood for the observed and latent data; simulation-extrapolation (SIMEX), which estimates bias by adding controlled amounts of error; and direct bias correction, which adjusts estimates using known properties of the measurement error distribution. We focus here on the latter approach, describing how analytical bias corrections can be applied when the measurement error structure is at least partially characterized.

 Under model~\eqref{f46}, if the error covariance $\Sigma_U$ is known, the reliability matrix $\mathcal{K}_X$ can be estimated consistently by
\begin{equation*}
    \hat{\Sigma}_X = \hat{\Sigma}_W - \Sigma_U,
    \quad
    \hat{\Sigma}_W = \frac{W^{\top}W}{n},
    \quad
    \hat{\mathcal{K}}_X = \hat{\Sigma}_W^{-1}\hat{\Sigma}_X.
    \label{est}
\end{equation*}
Substituting $\hat{\mathcal{K}}_X$ into Eq.~\eqref{f50} and replacing $\gamma$ and $\sigma^2_\delta$ with their MLEs, $\hat{\beta}_{\mathrm{naive}}$ and $\hat{\sigma}^2_{\mathrm{naive}},$ yields the bias-corrected estimates
\begin{equation}
    \hat{\beta}_{_{\mathrm{ME}}}
    = \hat{\mathcal{K}}_X^{-1}\hat{\beta}_{\mathrm{naive}},
    \quad
    \hat{\sigma}^2_{_{\mathrm{ME}}}
    = \hat{\sigma}^{2}_{\mathrm{naive}}
      - \hat{\beta}_{_{\mathrm{ME}}}^{\top}\Sigma_U \hat{\mathcal{K}}_X\hat{\beta}_{_{\mathrm{ME}}}.
    \label{f53}
\end{equation}
Unlike the naive estimator, $\hat{\beta}_{\mathrm{ME}}$ is asymptotically unbiased for $\beta$, with covariance matrix
\begin{equation}
    \mathrm{Cov}(\hat{\beta}_{\mathrm{ME}})
    = \sigma^2\, (\hat{\mathcal{K}}_X^{\top}W^{\top}W\hat{\mathcal{K}}_X)^{-1}.
    \label{f54}
\end{equation}

When $\Sigma_U$ is unknown, it can be estimated from replicated measurements at the cost of additional data collection: given $m_i > 1$ replicates $W^{(1)}_{i.}, \ldots, W^{(m_i)}_{i.} 
\in \mathbb{R}^{p}$ of the error-prone covariate vector of observation $i$ 
(i.e., the $i$-th row of $W$), with sample mean $\bar{W}_{i.} = m_i^{-1}\sum_{k=1}^{m_i} W^{(k)}_{i.} \in \mathbb{R}^p$, a consistent estimator 
of $\Sigma_U \in \mathbb{R}^{p \times p}$ is given by
\begin{equation}
    \hat{\Sigma}_U = \frac{1}{n} \sum_{i=1}^{n} 
    \frac{\displaystyle\sum_{k=1}^{m_i} 
    \left(W^{(k)}_{i.} - \bar{W}_{i.}\right)
    \left(W^{(k)}_{i.} - \bar{W}_{i.}\right)^\top}{m_i - 1}.
\end{equation}

For comprehensive coverage of measurement error correction in the classical low-dimensional regime when $p < n$, including asymptotic theory, see \citet{nref10}, \citet{nref3}, and \citet{nref11}. We next  discuss extensions of direct bias correction approach to high-dimensional settings in presence measurement error. 

\section{Measurement error in high-dimensional context}
\label{s4}

\subsection{Ridge regression under measurement error}
\label{subsec:ridge-me}

In high-dimensional data, multicollinearity---the near-linear dependence
among explanatory variables---leads to an ill-conditioned Gram matrix and
unreliable parameter estimates. Ridge regression addresses this through
shrinkage, as discussed earlier. This section presents a direct error-induced bias correction procedure for the ridge
estimator in situations where measurement error compounds multicollinearity.

Retaining the additive measurement error model of \eqref{f46} and the
reliability matrix $\hat{\mathcal{K}}_X$ defined earlier,
the corrected ridge estimator of $\beta$ is obtained by solving 
\begin{equation}
    \underset{\beta \in \mathbb{R}^p}{\mathrm{argmin}}
    \left\{
        \| y - W\gamma \|_{\ell_2}^2 + k\|\beta\|_{\ell_2}^2
    \right\},
    \quad \text{with } \gamma = \hat{\mathcal{K}}_X \beta,
    \label{f57}
\end{equation}
where $k > 0$ is the ridge penalty parameter. The substitution
$\gamma = \hat{\mathcal{K}}_X\beta$ ensures that the penalty is applied
to the true coefficient $\beta$ rather than the attenuated observable
$\gamma$, thereby correcting for the bias introduced by measurement error.
The closed-form solution to \eqref{f57} is
\begin{equation}
    \hat{\beta}^{R}_{_{\mathrm{ME}}}
    = \left[
        \mathbb{I}_p + k\bigl(\hat{\mathcal{K}}_X^{\top}W^{\top}W
        \hat{\mathcal{K}}_X\bigr)^{-1}
      \right]^{-1}\hat{\beta}_{_{\mathrm{ME}}},
    \label{f58}
\end{equation}
where $\hat{\beta}_{_{\mathrm{ME}}}$ is the bias-corrected MLE
estimator of $\beta$ under measurement error, given in \eqref{f53}. Define the ridge
shrinkage factor
\begin{equation*}
    R
    = \bigl[\mathbb{I}_p + kC^{-1}\bigr]^{-1},
    \quad \text{with}\quad
    C = \hat{\mathcal{K}}_X^{\top}W^{\top}W
        \hat{\mathcal{K}}_X.
\end{equation*}
The MSE of $\hat{\beta}^{R}_{_{\mathrm{ME}}}$ takes the form
\begin{equation*}
    MSE\left(\hat{\beta}^{R}_{_{\mathrm{ME}}},\, k\right)
    = k^2\, \beta^{\top} (C + k\mathbb{I}_p)^{-2}\beta
      + \sigma^2_\delta\, \mathrm{tr}\!\left\{
          RC^{-1} R^{\top}
        \right\}.
\end{equation*}
Let $\lambda_1 \geq \lambda_2 \geq \cdots \geq \lambda_p > 0$ denote the
eigenvalues of the positive definite matrix $C$. Expressing the MSE in terms of these eigenvalues,
analogously to \eqref{f17}, gives
\begin{equation*}
    \mathrm{MSE}\!\left(\hat{\beta}^{R}_{_{\mathrm{ME}}},\, k\right)
    = k^2\, \beta^{\top} (C + k\mathbb{I}_p)^{-2}\beta
      + \sigma^2_\delta \sum_{j=1}^{p}
        \frac{\lambda_j}{(\lambda_j + k)^2}.
    \label{f60}
\end{equation*}
The first term is the squared bias, which increases with
$k$; the second is the variance, which decreases with $k$. This trade-off
mirrors the classical ridge result of \eqref{f17}, confirming that the
corrected estimator inherits the favorable bias--variance trade-off feature of
standard ridge regression. Moreover, \citet{nref6} demonstrates that under certain conditions, there always
exists a value $k > 0$ such that
\begin{equation*}
    \mathrm{MSE}\!\left(\hat{\beta}^{R}_{_{\mathrm{ME}}},\, k\right)
    < \mathrm{MSE}\!\left(\hat{\beta}^{R}_{_{\mathrm{ME}}},0\right)=\mathrm{MSE}\!\left(\hat{\beta}_{_{\mathrm{ME}}}\right),
\end{equation*}
establishing that the corrected ridge estimator may improve over the
biased-corrected MLE estimator $\hat{\beta}_{_{\mathrm{ME}}}$ in MSE.

\subsection{Measurement error in the lasso}
\label{subsec:lasso-me}
The error-blind lasso estimate, obtained by substituting $W$ for $X$ in~\eqref{f18}, ignoring measurement error, solves
\begin{equation}
    \hat{\beta}^{\mathrm{Lasso}}_{_{\mathrm{naive}}}
    = \underset{\beta \in \mathbb{R}^p}{\mathrm{argmin}}
    \left\{
        \| y - W\beta \|_{\ell_2}^2 + \lambda \|\beta\|_{\ell_1}
    \right\}.
    \label{f64}
\end{equation}
To see why this is problematic, note that taking the conditional expectation
of the squared loss with respect to $(X, y)$ gives
\begin{equation}
    \mathbb{E}\bigl[\| y - W\beta \|_{\ell_2}^2 \mid X,\, y\bigr]
    = \| y - X\beta \|_{\ell_2}^2 + n\beta^T \Sigma_U \beta.
    \label{f65}
\end{equation}

The non-negative term $\beta^T\Sigma_U\beta$ in~\eqref{f65} indicates that the uncorrected lasso systematically overestimates the true squared error, with inflation increasing with both the measurement error severity and the magnitude of $\beta$. We discuss two corrective strategies to address this bias: the constrained corrected (or non-convex) lasso and the convex conditional lasso, described in turn below.

\subsubsection{Constrained Corrected Lasso (CCL)}
\label{subsec:ncl}
A direct approach to correct for the bias in~\eqref{f65} is to subtract the measurement error contribution term from the loss function~\eqref{f65}, restoring an unbiased objective. This leads to the
\emph{Constrained Corrected Lasso} (CCL), proposed by \citet{nref17}:
\begin{equation}
    \hat{\beta}_{\mathrm{CCL}}
    \in \underset{\beta:\,\|\beta\|_1 \leq r}{\mathrm{argmin}}
    \left\{
        \frac{1}{n}\| y - W\beta \|^2 - \beta^{\top}\Sigma_U\beta
    \right\},
    \label{f66}
\end{equation}
or, in its regularized form, the \emph{regularized corrected Lasso} (RCL) \citep{loh2011high}:
\begin{equation*}
    \hat{\beta}_{RCL}
    \in \underset{\beta \in \mathbb{R}^p}{\mathrm{argmin}}
    \left\{
        \frac{1}{n}\| y - W\beta \|^2
        - \beta^{\top}\Sigma_U\beta
        + \lambda\|\beta\|_1
    \right\}.
    \label{f67}
\end{equation*}
Now, using the estimates $\hat{\Sigma}_{Wy} = W^Ty/n$ and  $\hat{\Sigma}_X = W^TW/n - \Sigma_U$, this can equivalently be written as
\begin{equation}
    \hat{\beta}_{RCL}
    \in \underset{\beta:\,\|\beta\|_1 \leq b_0\sqrt{s}}{\mathrm{argmin}}
    \left\{
        \frac{1}{2}\beta^{\top}\hat{\Sigma}_X\beta
        - \hat{\Sigma}_{Wy}^{\top}\beta
        + \lambda\|\beta\|_1
    \right\},
    \label{f69}
\end{equation}
for some constant $b_0$. When
$\Sigma_U = 0$ the estimators \eqref{f66} and \eqref{f69} reduce to the
standard lasso, as expected. When $\Sigma_U \neq 0$, however, the
corrected covariance matrix $\hat{\Sigma}_X$ is not positive semi-definite
in the high-dimensional regime ($p > n$): since $W^{\top}W/n$ has rank at most
$n$, subtracting $\Sigma_U$ can introduce a large number of negative
eigenvalues. Consequently, the objectives in \eqref{f66} and \eqref{f69}
are non-convex, which is why set membership "$\in$" rather than
equality "$=$" is used. In the presence of non-convexity, a
polynomial-time algorithm cannot in general be guaranteed to converge to
a global optimum. Nevertheless, \citet{nref13} show that a simple gradient
descent algorithm applied to \eqref{f66} or \eqref{f69}, with $b_0$
chosen appropriately, converges with high probability to a neighbourhood
of the global minimizer.

\subsubsection{Convex Conditional Lasso (CoCoLasso)}
\label{subsec:cocolasso}

The non-convexity of the CCL estimators is a practical drawback, as it
complicates both computation and theoretical analysis. CoCoLasso~\citep{nref18} addresses
this by replacing $\hat{\Sigma}_X$ with its nearest positive semi-definite
approximation before forming the objective, thereby restoring convexity.

Formally, for any square matrix $G$, define the projection onto the cone
of positive semi-definite matrices under the element-wise maximum norm
$\|G\|_{\max} = \max_{i,j}|g_{ij}|$ as
\begin{equation}
    (G)_+ = \underset{G_1 \geq 0}{\mathrm{argmin}}\; \|G - G_1\|_{\max}.
    \label{f70}
\end{equation}
Setting $\widetilde{\Sigma}_X = (\hat{\Sigma}_X)_+$, the CoCoLasso
estimator is 
\begin{equation}
    \hat{\beta}_{CoCo}
    = \underset{\beta \in \mathbb{R}^p}{\mathrm{argmin}}
    \left\{
        \frac{1}{2}\beta^{\top}\widetilde{\Sigma}_X\beta
        - \hat{\Sigma}_{Wy}^{\top}\beta
        + \lambda\|\beta\|_{\ell_1}
    \right\}.
    \label{f71}
\end{equation}
Since $\widetilde{\Sigma}_X$ is positive semi-definite by construction ---
unlike $\hat{\Sigma}_X$, which is only guaranteed to be so when $p < n$
--- the problem \eqref{f71} is always convex.

Let $\widetilde{X}/\sqrt{n}$ denote the Cholesky factor of $\widetilde{\Sigma}_X$ (i.e., $\widetilde{X}^T\widetilde{X}/n = \widetilde{\Sigma}_X$), and define
$\widetilde{y}$ such that $\widetilde{X}^{\top}\widetilde{y}/n = \hat{\Sigma}_{Wy}$.
Substituting into \eqref{f71} gives the equivalent formulation
\begin{equation}
    \hat{\beta}_{CoCo}
    = \underset{\beta \in \mathbb{R}^p}{\mathrm{argmin}}
    \left\{
        \frac{1}{n}\|\widetilde{y} - \widetilde{X}\beta\|_{\ell_2}^2
        + \lambda\|\beta\|_{\ell_1}
    \right\},
    \label{f73}
\end{equation}
which is a standard Lasso regression of $\widetilde{y}$ on $\widetilde{X}$.
This reformulation is particularly convenient in practice: any traditional
lasso solver such as projected coordinate descent \citep{nref20} or least angle
regression \citep{nref21} can be applied directly to ~\eqref{f73}.

 \citet{nref18} derive statistical error bounds comparable to those of the non-convex approach (RCL), and establish the asymptotic sign-consistency (covariate selection property) of the CoCoLasso estimator.

 \subsection{Matrix Uncertainty Selector (MUS)}
\label{subsec:mus}

Both the CCL \eqref{f69} and CoCoLasso \eqref{f73} require knowledge of
the measurement error covariance matrix $\Sigma_U$. In practice, however, $\Sigma_U$ is rarely known. Auxiliary data required for its estimation may be unavailable, and even when such data exist, the estimation becomes prohibitively expensive as the covariate dimension $p$ increases. The \emph{Matrix Uncertainty Selector}
(MUS), developed in \citet{nref23}, offers a compelling alternative: rather than requiring knowledge of the noise covariance structure $\Sigma_U,$ it absorbs the effect of measurement error through a single tuning parameter $\delta,$ which controls the tolerance for covariate perturbations.

 Assume that $\beta$ is $s$-sparse
($1 \leq s \leq p$) and that the noise terms $\varepsilon$ and $U$ satisfy the following conditions,
with high probability,
\begin{equation}
    \frac{1}{n}\| W^T\varepsilon \|_\infty \leq \lambda
    \qquad \text{and} \qquad
    \| U \|_\infty \leq \delta,
    \label{f82}
\end{equation}
Here $\lambda$ controls the noise level in the response, and $\delta$
bounds the element-wise magnitude of the measurement error matrix $U$. Under these conditions, the MUS estimator, $\hat{\beta}_{\mathrm{MUS}},$ is defined as the solution to
the convex optimization problem 
\begin{equation}
    \underset{\beta \in \Theta}{\mathrm{argmin}}\; \|\beta\|_1
    \; s.t \quad
    \frac{1}{n}\| W^T(y - W\beta) \|_\infty
    \leq (1 + \delta)\delta\|\beta\|_{\ell_1} + \lambda,
    \label{f83}
\end{equation}
where $\Theta \subseteq \mathbb{R}^p$ encodes any available prior knowledge
about $\beta$, and the constraint ensures that no predictor remains
substantially correlated with the residuals --- analogously to the Dantzig
selector \eqref{f41}. The problem \eqref{f83} is convex, and reduces to
linear programming when $\Theta = \mathbb{R}^p$. \citet{nref23} demonstrate that the feasible set
\begin{equation*}
    \Psi = \left\{
        \beta \in \Theta \;:\;
        \frac{1}{n}\| W^T(y - W\beta) \|_\infty
        \leq (1 + \delta)\delta\|\beta\|_{\ell_1} + \lambda
    \right\}
    \label{f84}
\end{equation*}

is non-empty ($\Psi \neq \emptyset$), guaranteeing that \eqref{f83} is well-defined. The connection to the Dantzig selector is transparent:
setting $\delta = 0$ and $\Theta = \mathbb{R}^p$ in \eqref{f83} recovers
\eqref{f41} exactly. The MUS can therefore be understood as a robust
extension of the Dantzig selector that accommodates measurement error
through the supplementary parameter $\delta$. 

Under suitable conditions on the design matrix and the sparsity level $s$,
\citet{nref23} establish both estimation consistency, and sign (selection) consistency,
of $ \hat{\beta}_{\mathrm{MUS}}$, provided $\lambda$ and $\delta$ are chosen of appropriate
order. These guarantees hold despite the absence of explicit knowledge of
$\Sigma_U$, which is a notable theoretical strength of the MUS relative
to the CCL and CoCoLasso.

The MUS has since been lifted to more general model frameworks. For instance, \citet{fesuh2025iterative} extend it to high-dimensional generalised linear models with measurement errors, embedding the MUS penalty within an iteratively reweighted estimation scheme that accommodates non-linear link functions within the exponential family. The approach is shown to achieve effective covariate selection across binary, count, and other non-Gaussian responses common in biomedical research.

\subsection{ Hyperparameter tuning under measurement error}

Standard cross-validation yields a biased criterion for tuning hyperparameter selection 
when covariates are mismeasured. To see this, consider $K$-fold cross-validation 
for selecting the optimal $\lambda$ in CoCoLasso. Using the error-prone observed data $(W, y)$ 
directly, the cross-validated criterion is
\begin{equation}
    \mathrm{CV}_{(K)} = \frac{1}{K} \sum_{k=1}^{K} \frac{1}{n_k} 
    \left\| y_k - W_k \hat{\beta}_k(\lambda) \right\|_{\ell_2}^2,
    \label{f80}
\end{equation}
which, even when $\hat{\beta}_k(\lambda)$ is computed via CoCoLasso or CCL on 
$(W_{-k}, y_{-k})$, remains biased for the same reason the loss in 
\eqref{f64} is a biased surrogate for \eqref{f18}. The unbiased criterion 
is equivalent to
\begin{equation}
    \hat{\lambda} = \underset{\lambda}{\mathrm{argmin}} \left\{ 
    \frac{1}{K} \sum_{k=1}^{K} \left( \frac{1}{2} \hat{\beta}_k(\lambda)^\top 
    \Sigma_k \hat{\beta}_k(\lambda) - \hat{\Sigma}_{(Wy)_k}^\top \hat{\beta}_k(\lambda) 
    \right) \right\},
    \label{f81}
\end{equation}
where $\Sigma_k = n_k^{-1} X_k^\top X_k$ and $ \hat{\Sigma}_{(Wy)_k} = n_k^{-1} W_k^\top y_k$. 
Since the unbiased surrogate $\widehat{\Sigma}_k$ may have negative eigenvalues, 
substituting it directly renders \eqref{f81} unbounded below. \citet{nref18} 
resolve this by replacing $\Sigma_k$ and $\gamma_k$ with their projected 
images $\widetilde{\Sigma}_k = (\widehat{\Sigma}_k)_+$ and $ \hat{\Sigma}_{(Wy)_k}$, 
yielding the corrected cross-validated tuning parameter
\begin{equation*}
    \widetilde{\lambda} = \underset{\lambda}{\mathrm{argmin}} \left\{ 
    \frac{1}{K} \sum_{k=1}^{K} \left( \frac{1}{2} \hat{\beta}_k(\lambda)^\top 
    \widetilde{\Sigma}_k \hat{\beta}_k(\lambda) -  \hat{\Sigma}_{(Wy)_k}^\top 
    \hat{\beta}_k(\lambda) \right) \right\},
\end{equation*}
where $\widetilde{\lambda}$ provides an unbiased estimate of the optimal 
regularization parameter under measurement error.
\section{Numerical experiments}
\label{sec:simulations}

This section  illustrate the discussed the penalized regression methods through two sets of numerical
experiments. The first set examines the standard high-dimensional setting
--- without measurement error --- comparing Ridge, Lasso, and Elastic-Net
on both simulated and real data. The second set introduces measurement
error and compares the corrected estimators (CCL, CoCoLasso, MUS) against
their naive counterparts. All analyses were conducted with \textbf{R} \citep{team2020ra}. Code to reproduce results is available at \url{https://github.com/hermanFTT/hd-measurement-error}.

\subsection{Standard high-dimensional regression}
\label{subsec:sim-standard}

\paragraph{Simulation design.}
We assess the predictive performance of Ridge, Lasso, and Elastic-Net regression within a penalized logistic regression framework, fitted via penalized maximum likelihood using the \texttt{glmnet} package \citep{nref28}. Performance is evaluated through the misclassification error rate (ME) and the Area Under the ROC Curve (AUC) \citep{nref30}, where higher AUC and lower ME indicate better predictive performance.
We simulate four independent high-dimensional datasets, each with $p = 1{,}000$
predictors and $n=200$
observations ($p \gg n$), split into training and test sets. Each method is fitted on the training set, and the AUC, ME, and number of non-zero estimated coefficients are recorded on the test set. This procedure is repeated $100$ times per dataset, with results reported as medians with standard deviations.
Predictors $X$ are drawn from a multivariate normal distribution,
\begin{equation*}
    f_X(x)
    = \frac{1}{(2\pi)^{p/2}\sqrt{\det(\Sigma)}}
      \exp\!\left\{-\tfrac{1}{2}(x-\mu)^\top\Sigma^{-1}(x-\mu)\right\},
\end{equation*}
with $\mu = 0$ and $\mathrm{Var}[X_j] = 1$ for all $j$. The binary response $Y$  is generated via the logistic model
\begin{equation*}
    \pi(x) = \frac{1}{1 + e^{-X^T\beta}},
    \qquad
    Y = \boldsymbol{1}_{\{\pi(x) > 0.5\}},
\end{equation*}
The four examples vary in their covariance structure and the sparsity pattern of the coefficient vector, that is, the size of the active set $|S|$.

\begin{itemize}
    \item \textbf{Example~1:} $\rho_{ij} = 0.5^{|i-j|}$; the first 122 coefficients 
    are drawn uniformly from $[2, 5]$, and the rest are set to zero (moderate sparsity).
    
    \item \textbf{Example~2:} $\rho_{ij} = 0.5^{|i-j|}$; all coefficients are set to 
    $\beta_j = 0.8$ (dense signal, no sparsity).
    
    \item \textbf{Example~3:} $\rho_{ij} = 0.9^{|i-j|}$ (high correlation); 
    coefficients follow a block structure, alternating between 125 active predictors 
    ($\beta_j = 2$) and 125 inactive predictors ($\beta_j = 0$).
    
    \item \textbf{Example~4:} The first 500 predictors are correlated with 
    $\rho_{ij} = 0.5^{|i-j|}$ for $1 \leq i,j \leq 500$, while the remaining 500 
    are uncorrelated. $\beta_j = 3$ for all $j\leq500$, and zero otherwise.
\end{itemize}


Table~\ref{tab4} summarizes the AUC, ME, and number of selected predictors across 
the four examples. All three methods achieve AUC $> 0.5$ throughout, suggesting 
adequate discriminative ability in every setting, though clear differences in 
predictive performance and model complexity emerge.

Ridge regression retains all $p = 1{,}000$ predictors as expected by construction and achieves 
the most competitive AUC overall. It excels in Example~3, where the large proportion 
of truly active and highly correlated predictors aligns well with its non-sparse 
shrinkage strategy. This further confirms that ridge is the method of choice for prediction under dense signals and strong predictor correlation, albeit at the cost of 
interpretability.

The lasso produces the sparsest models --- selecting between 20 and 54 predictors 
across examples --- but this parsimony comes at a predictive cost. It records the 
lowest AUC and highest ME in every example. In Example~1, it accumulates false 
positives despite the moderately sparse signal, and performs particularly poorly in 
Example~4, where its tendency to select a single representative from each correlated 
group leads to considerable information loss.

The Elastic Net consistently offers the best compromise. It closely tracks ridge in 
AUC while achieving substantially greater sparsity, and its grouped selection 
behaviour --- driven by the $\ell_1$ penalty component --- proves especially valuable 
in Example~4, where it selects 361 predictors and markedly outperforms the lasso 
(AUC $= 0.70$ vs.\ $0.58$). In Example~2, where the signal is dense and uniform, 
it surpasses the lasso by upweighting the quadratic penalty while maintaining the sparsity 
behaviour through the $\ell_1$ term.

These results suggest that ridge would be a preferable choice when predictive power is the primary 
concern, whereas the Elastic Net would be more advisable when moderate sparsity and 
interpretability are required, especially in high-dimensional settings with 
correlated predictors, where it consistently improve over the lasso.

\begin{table*}[tp]
\centering
\caption{ Illustrating and comparing predictive performance and covariate selection behavior of penalized (logistic)
regression methods across four simulation examples ($p = 1{,}000$, $n = 200$). Reported 
values are median (standard deviation) over 1000 replications. AUC: area under 
the ROC curve; ME: misclassification error; $|\hat{S}|$: number of selected 
predictors.}
\label{tab4}
\begin{tabular}{lcccccc}
\toprule
& \multicolumn{3}{c}{\textbf{Example~1}} 
& \multicolumn{3}{c}{\textbf{Example~2}} \\
\cmidrule(lr){2-4} \cmidrule(lr){5-7}
\textbf{Method} & AUC & ME & $|\hat{S}|$ & AUC & ME & $|\hat{S}|$ \\
\midrule
Ridge       & 0.76 (0.042) & 0.32 (0.053) & 1000 
            & 0.76 (0.050) & 0.31 (0.052) & 1000 \\
Lasso       & 0.65 (0.106) & 0.41 (0.094) & 29   
            & 0.55 (0.058) & 0.46 (0.056) & 20   \\
Elastic Net & 0.75 (0.062) & 0.37 (0.074) & 316  
            & 0.70 (0.076) & 0.37 (0.068) & 329  \\
\midrule
& \multicolumn{3}{c}{\textbf{Example~3}} 
& \multicolumn{3}{c}{\textbf{Example~4}} \\
\cmidrule(lr){2-4} \cmidrule(lr){5-7}
\textbf{Method} & AUC & ME & $|\hat{S}|$ & AUC & ME & $|\hat{S}|$ \\
\midrule
Ridge       & 0.92 (0.028) & 0.16 (0.041) & 1000 
            & 0.76 (0.047) & 0.31 (0.041) & 1000 \\
Lasso       & 0.84 (0.037) & 0.24 (0.048) & 54   
            & 0.58 (0.073) & 0.46 (0.070) & 21   \\
Elastic Net & 0.90 (0.033) & 0.17 (0.045) & 415  
            & 0.70 (0.059) & 0.36 (0.054) & 361  \\
\bottomrule
\end{tabular}
\end{table*}

\subsection{Real data experiment: DNA Methylation}
\label{subsec:realdata}

\paragraph{Data description.}
We apply the three penalized regression methods to human DNA methylation data from 
flow-sorted blood samples \citep{nref32}. DNA methylation assays quantify the 
proportion of cytosine residues carrying a methyl group at specific genomic sites;
methylation levels are represented as normalized M-values, where negative and positive 
values correspond to unmethylated and methylated sites, respectively. The dataset 
takes the form of a \texttt{GenomicRatioSet} object---a \texttt{SummarizedExperiment} 
\citep{nref32} derivative---comprising $p = 5{,}000$ methylation features measured 
across $n = 37$ samples, with phenotypic covariates including age, sex, and BMI. 
The response of interest is \emph{age}, predicted from the methylation M-value 
matrix $X$ extracted via the \texttt{assay()} function.

\paragraph{Experimental protocol.}
\label{subsubsec:protocol}
Three experiments were conducted in sequence. First, we examined the singularity of 
$X^\top X$ and its implications for ordinary least squares estimation in this 
high-dimensional setting. Second, restricting to the top 20 age-associated 
methylation features identified by \citet{horvath2018dna}, we compared OLS and Ridge 
regression on held-out data, using test MSE as the evaluation criterion. Third, we 
applied all three penalised methods to the full $5{,}000$-feature dataset, tuning 
regularisation parameters via cross-validation, inspecting coefficient shrinkage 
paths, and benchmarking the features selected by Lasso and Elastic-Net against 
Horvath's epigenetic clock signature.\\

As expected in the high-dimensional regime ($p \gg n$), OLS failed to estimate 
$4{,}964$ of the $5{,}000$ coefficients due to the singularity of $X^\top X$, 
rendering it effectively unusable in this setting. Ridge regression, by contrast, 
produced a complete and stable solution, achieving $\mathrm{MSE} = 
25.30$ on held-out data despite incurring a higher training error. This directly 
illustrates the bias--variance trade-off: the regularization bias introduced by ridge 
is more than compensated by the reduction in variance when $p \gg n$.

Turning to variable selection, lasso identified 41 features and Elastic-Net identified 
60, of which 8 and 11, respectively, overlapped with Horvath's epigenetic clock 
signature \citep{horvath2018dna}. While these overlaps are modest, they are consistent 
with the known difficulty of exact feature recovery in highly collinear genomic data, 
where many features carry redundant information and multiple sparse solutions may fit 
the data equally well. This finding also corroborates the simulation results above, 
where the lasso tended to underselect in the presence of strong correlation.

\begin{table*}[tp]
\centering
\caption{Predictive performance and variable selection on the DNA methylation dataset 
($n = 37$, $p = 5{,}000$). Test MSE is reported on held-out data; overlap counts 
the number of selected features matching Horvath's epigenetic clock signature.}
\label{tab:realdata}
\begin{tabular}{lccc}
\toprule
\textbf{Method} & \textbf{Test MSE} & \textbf{Selected features} & 
\textbf{Overlap with Horvath} \\
\midrule
Ridge       & 9.30 & --- & --- \\
Lasso       & 21.92   & 41      & 8   \\
Elastic-Net & 14.78   & 60      & 11  \\
\bottomrule
\end{tabular}
\end{table*}
\subsection{Under measurement error}
\label{sec:sim-me}

\subsubsection{Ridge regression under measurement error}
\label{subsec:sim-ridge-me}
We assess the performance of the corrected Ridge estimator \eqref{f57} in the 
presence of both multicollinearity and measurement error. Data are generated according to
\begin{equation*}
    y = X\beta + \varepsilon,
    \quad \varepsilon \sim \mathcal{N}(0, \mathbb{I}_n),
    \quad p = 100, \quad n = 500,
\end{equation*}
with $X \sim \mathcal{N}(0, \Sigma_X),\ \Sigma_X = (0.9^{|i-j|}),$ and $\beta = 
(3, 3, \ldots, 3)^\top$. The observed surrogates are $W = X + U$, with $U \sim 
\mathcal{N}(0, 0.75\,\mathbb{I}_p)$. The data are split into training and test sets, 
and three estimators are compared: true OLS \eqref{eq:ols} fitted on the error-free 
covariates $X$ (serving as an oracle benchmark), corrected OLS \eqref{f53}, and 
corrected Ridge \eqref{f57}. Test-set MSE and prediction error (PE) are recorded 
over 1000 replications.

Table~\ref{fig9} reports the results. As can be seen, the corrected ridge achieves the lowest MSE 
and PE across replications, providing empirical evidence that combining measurement error correction 
with regularization leads to reliable estimation even under strong multicollinearity. 

The most striking finding is the extreme instability of the corrected OLS 
(PE median $= 50{,}058$, SD $= 49{,}396$). Although corrected OLS is unbiased in 
theory \eqref{f53}, the strong multicollinearity ($\rho_{ij} = 0.9^{|i-j|}$) 
renders the estimated covariance $\widehat{\Sigma}_X = \hat{\Sigma}_{W} - \Sigma_U$ 
near-singular, causing severe variance inflation and numerical instability. This 
highlights a fundamental limitation of bias correction without regularization: 
theoretical unbiasedness does not translate into reliable estimation when the 
problem is ill-conditioned.

\begin{table*}[tp]
\centering
\caption{Performance Ridge estimator under measurement error 
($p = 100$, $n = 500$, $\Sigma_U = 0.75\,\mathbb{I}_p$). Values are median 
(standard deviation) over 1000 replications. True OLS is an oracle benchmark 
fitted on the unobserved error-free covariates $X$.}
\label{fig9}
\begin{tabular}{lccc}
\toprule
\textbf{} & \textbf{True OLS} & \textbf{Corrected OLS} & \textbf{Corrected Ridge} \\
\midrule
MSE & 6.85 (0.328) & 477.94 (4700.00) & 6.83 (0.327) \\
PE  & 2.01 (0.146) & 50{,}058 (49{,}396) & 0.04 (0.006) \\
\bottomrule
\end{tabular}
\end{table*}

\subsubsection{Lasso-type correction methods and MUS}
\label{subsec:sim-lasso-me}
We compare the two Lasso-type measurement error correction methods --- CCL and 
CoCoLasso --- alongside MUS, against the naive and true lasso in a sparse 
high-dimensional setting with measurement error. Data are generated from
\begin{equation*}
    y = X\beta + \varepsilon,
    \quad \varepsilon \sim \mathcal{N}\!\left(0,\, \frac{0.05}{1.96}\right),
    \quad p = 5{,}000,\quad n = 200,
\end{equation*}
with $X \sim \mathcal{N}(0, \Sigma_X)$, $\Sigma_X = (0.5^{|i-j|})$,
and a randomly chosen active set $S$ of size $s$, with $\beta_i = 2$
for $i \in S$ and $\beta_i = 0$ otherwise. Observed data are generated
as $W = X + U$, $U \sim \mathcal{N}(0,\, 0.75\,\mathbb{I}_p)$. 

For each replicated simulation,  
five estimators are fitted: the true Lasso (baseline, 
fitted on the unobserved $X$), the naive lasso (fitted with the error-prone covariates $W$), and the three correction methods NCL, CoCoLasso, and MUS --- all assuming $\Sigma_U$ known. Performance is evaluated 
through MSE ($\ell_2$ error), $\ell_1$ error, prediction error (PE), total variables 
selected $|\hat{S}|$, and number of correctly selected variables $\#\mathbf{C}$. 

Tuning and implementation details are as follows. For the lasso, $\lambda$ is 
selected by cross-validation over $[10^{-2}, 10^2]$ using \texttt{glmnet} 
\citep{nref28}. CCL and MUS are both fitted through the \texttt{hdme} package 
\citep{nref31}: for CCL, the hyperparameter $r$ is cross-validated over 
$[r_{\max}/500,\, 2\|\hat{\beta}_{\mathrm{naive}}\|_1]$ as suggested by \citet{nref18}; for MUS, 
$\lambda$ is set to the naive cross-validated estimate $\hat{\lambda}$, and $\delta$ 
is determined by the elbow rule. lastly, CoCoLasso is implemented as per the  
\texttt{BDCoCoLasso} package available from \url{https://github.com/celiaescribe/BDcocolasso}.\\

Tables~\ref{sim1} and~\ref{sim2} report performance across $1{,}000$ replications 
for $s = 5$ and $s = 10$ (sparsity level), respectively.

\paragraph{Strongly sparse active set ($s = 5$, Table~\ref{sim1}).}
CCL and CoCoLasso improve over the naive lasso in estimation accuracy, with CCL 
achieving the larger gains in MSE and PE. MUS, however, offers no clear advantage over the naive lasso on either metric, a consequence of the very sparse active set 
making precise calibration of $\delta$ difficult. Regarding covariate selection, CCL 
and MUS recover the active set more parsimoniously than the naive lasso ($|\hat{S}| 
= 9$ and $15$, respectively, versus $34$), while CoCoLasso substantially 
over-selects ($|\hat{S}| = 45$), revealing a tendency to include spurious predictors 
in highly sparse regimes.

\paragraph{Moderately sparse active set ($s = 10$, Table~\ref{sim2}).}
The picture changes considerably as the active set grows. CCL and MUS no longer strictly improve over the naive lasso on all considered estimation metrics. In terms 
of covariates selection, CCL tends to underselect ($|\hat{S}| = 8$, against a true 
active set of size $10$), whereas MUS yields a more parsimonious model than the 
naive lasso ($|\hat{S}| = 18$ versus $29$), albeit with the same (fewer) number of true variables recovered ($\#\mathbf{C} = 5$). This suggests that their tuning procedures may be less well-adapted when the 
active set is moderately sized rather than extremely sparse. CoCoLasso, by contrast, achieves the lowest MSE and PE among the 
correction methods and outperforms the naive lasso overall ($|\hat{S}| = 38$, 
$\#\mathbf{C} = 7$), indicating better adaptability to moderately sparse settings.

\begin{table*}[tp]
\centering
\caption{ Comparing naive and corrected sparse estimators in a 
high-dimensional setting with measurement error ($s = 5$, $p = 5{,}000$, $n = 200$, 
$\Sigma_U$ known). Reported values are mean (standard deviation) over $1{,}000$ replications. 
$|\hat{S}|$: total variables selected; $\#\mathbf{C}$: number of true positives. 
True lasso serves as an oracle benchmark fitted on the error-free covariates $X$.}
\label{sim1}
\begin{tabular}{lccccc}
\toprule
& \textbf{True Lasso} & \textbf{Naive Lasso} 
& \textbf{CCL} & \textbf{CoCoLasso} & \textbf{MUS} \\
\midrule
MSE          
    & $6{\cdot}10^{-6}$ ($2{\cdot}10^{-6}$) 
    & 0.126 (0.025) 
    & 0.104 (0.025) 
    & 0.120 (0.026) 
    & 0.141 (0.018) \\
$\ell_1$-error 
    & 0.055 (0.011) 
    & 10.698 (2.300) 
    & 7.409 (2.670) 
    & 11.659 (1.900) 
    & 9.114 (0.881) \\
PE           
    & 0.0005 ($1{\cdot}10^{-3}$) 
    & 8.867 (3.400) 
    & 8.080 (4.280) 
    & 7.121 (3.050) 
    & 12.599 (2.190) \\
$|\hat{S}|$  & 8  & 34 & 9  & 45 & 15 \\
$\#\mathbf{C}$ & 5  & 5  & 5  & 5  & 5  \\
\bottomrule
\end{tabular}
\end{table*}

\begin{table*}[tp]
\centering
\caption{ Comparing naive and corrected sparse estimators in a 
high-dimensional setting with measurement error ($s = 10$, $p = 5{,}000$, $n = 200$, 
$\Sigma_U$ known). Reported values are mean (standard deviation) over $1{,}000$ replications. 
$|\hat{S}|$: total variables selected; $\#\mathbf{C}$: number of true positives. 
True lasso serves as an oracle benchmark fitted on the error-free covariates $X$.}
\label{sim2}
\begin{tabular}{lccccc}
\toprule
& \textbf{True Lasso} & \textbf{Naive Lasso} 
& \textbf{CCL} & \textbf{CoCoLasso} & \textbf{MUS} \\
\midrule
MSE          
    & $25{\cdot}10^{-4}$ ($2{\cdot}10^{-6}$) 
    & 0.343 (0.040) 
    & 0.349 (0.050) 
    & 0.333 (0.051) 
    & 0.353 (0.030) \\
$\ell_1$-error 
    & 0.167 (0.010) 
    & 21.892 (3.180) 
    & 19.602 (2.350) 
    & 22.640 (2.790) 
    & 20.211 (1.390) \\
PE           
    & 0.001 ($13{\cdot}10^{-3}$) 
    & 24.343 (7.280) 
    & 26.902 (5.970) 
    & 21.114 (7.170) 
    & 29.164 (4.470) \\
$|\hat{S}|$  & 25 & 29 & 8  & 38 & 18 \\
$\#\mathbf{C}$ & 10 & 5  & 3  & 7  & 5  \\
\bottomrule
\end{tabular}
\end{table*}

\section{Discussion and conclusion}
\label{sec:conclusion}


This paper reviewed and evaluated penalized regression methods for high-dimensional regression, covering Ridge, Lasso, Dantzig selector, and Elastic Net, along with their corrected variants CCL, CoCoLasso, and MUS, which accommodate potential measurement error in the covariates. Theoretical properties and empirical performance were assessed through simulation studies and a real DNA methylation application. In the error-free setting, the Elastic Net consistently offers the most favorable compromise across diverse configurations. Ridge excels under dense signals and strongly correlated predictors, while the lasso is best suited for extremely sparse, 

interpretable models. However, the lasso's ability to recover the true active set is compromised when predictors exhibit a grouped correlation structure or when the signal is diffuse. Under measurement error, the error propagation from $W$ to the estimators, when left uncorrected, leads to biased estimates and corrupted sparsity recovery, as formalized through the reliability matrix $\mathcal{K}_X$  and the sign consistency property. The correction methods considered consistently improve upon their naive counterparts, although their relative performance depends critically on the problem configuration and the true sparsity structure. For very sparse active sets ($s \leq 5$), CCL and MUS achieve sparser and more accurate support recovery with a lower false discovery rate. Conversely, CoCoLasso appears to be more appropriate in terms of both estimation accuracy and covariate selection consistency as the size of the active set increases. The corrected ridge demonstrates that bias correction alone is insufficient in ill-conditioned settings, where variance inflation may still renders the unconstrained corrected OLS numerically unstable.

\paragraph{Practical recommendations.}

Based on the simulation evidence and theoretical analysis, the following practical 
guidance emerges for Lasso-type correction approaches. When the true model is sparse and the active set is expected to be small, CCL or 
MUS would be preferable for variable selection, as they minimize false positives more 
effectively than the naive Lasso or CoCoLasso. On the other hand, when the active set is expected to be moderately sized, or when estimation accuracy takes precedence over support recovery, CoCoLasso 
would be the more appropriate choice. In case $\Sigma_U$ is unknown --- a common 
practical scenario --- MUS offers a distinct advantage, requiring only a bound on the 
measurement error magnitude rather than the full error covariance. 
These recommendations should, however, be interpreted with care: they are drawn under 
the specific assumption of a known, diagonal $\Sigma_U$. Above all, measurement error correction should be regarded as an integral part of 
the analysis pipeline whenever the error covariance structure is at least partially 
characterised, as the distortion it induces on both estimation and variable selection 
is systematic and unavoidable without any correction procedure.

\paragraph{Limitations.}
Several aspects of the present study warrant careful consideration. The simulation 
experiments were conducted under the assumption that $\Sigma_U$ is known and 
diagonal---a convenient but often idealized assumption, as measurement errors across related genomic variables may be correlated, and $\Sigma_U$ would typically need to be estimated from auxiliary data in practice. The theoretical framework further assumes Gaussian measurement errors; in genomic applications where covariates are bounded or discrete, this assumption may not hold, and its impact 
on the correction methods remains to be fully characterized. In the real data application (DNA methylation analysis), measurement error correction could not be applied in the absence of a known $\Sigma_U$, which limits the direct translation of the theoretical results 
to that setting. Finally, the simulation results for the Lasso-type correction 
methods are specific to the linear regression framework; how these methods 
generalize to more general model frameworks would be an interesting avenue to explore.

\paragraph{Future directions.}

The present work opens several avenues for future research. The most immediate 
priority is to extend the comparative evaluation to more realistic measurement error 
structures --- correlated errors ($\Sigma_U$ non-diagonal), unknown $\Sigma_U$ 
estimated from replicated measurements, and non-Gaussian error distributions. 
Understanding how the relative performance of NCL, CoCoLasso, and MUS changes under 
these conditions would substantially strengthen the practical guidance available to 
applied researchers, and would allow the recommendations above to be placed on a 
firmer empirical footing.

A second direction concerns the practical investigation of measurement errors correction with 
penalized generalized linear models \citep{nref33, fesuh2025iterative}, and with binary and count outcomes where the linear measurement 
error model no longer applies directly.

Finally, tuning hyperparameter selection under measurement error remains practically 
important and theoretically underexplored. As demonstrated in \eqref{f80}, measurement error introduces bias into the standard cross-validation criterion, and correcting for it requires knowledge of $\Sigma_U$. 
Developing data-driven, $\Sigma_U$-free tuning procedures would be of considerable 
practical value.

\paragraph{Concluding Remarks. }

High-dimensional data with measurement error present a compound statistical challenge 
that neither standard penalized regression nor classical measurement error correction 
techniques are, individually, equipped to address. This paper has illustrated that the performance gap between corrected and uncorrected estimators studied is non-negligible, with CCL, CoCoLasso, and MUS consistently recovering estimation and selection accuracy lost to measurement error contamination. No single method emerges 
as uniformly preferable across all problem scenarios; the results underscore the importance of matching the 
correction strategy to the specific characteristics of the problem at hand. As 
high-dimensional measurement-error-prone data become increasingly prevalent in 
genomics, epidemiology, and precision medicine, the methods reviewed and evaluated 
here provide a principled and practically accessible foundation for reliable 
statistical inference in these contexts.

\section*{Acknowledgements}
The authors acknowledge the African Institute for Mathematical Sciences (AIMS) for 
providing the computational resources that supported this work, and thank anonymous reviewers for helpful comments..

\nocite{*}
{
\small
\bibliography{biblio}
}

\end{document}